\documentclass[11pt]{article}
\usepackage[normalem]{ulem}   %%%%%  PACKAGE TO CROOS OUT

 \RequirePackage{graphicx}
\RequirePackage[colorlinks,citecolor=blue,urlcolor=blue,linkcolor=blue]{hyperref}

\usepackage[utf8]{inputenc}
\usepackage[english]{babel}

\usepackage{amsmath}
\usepackage{amsfonts}
\usepackage{amssymb}
\usepackage{graphicx}
\usepackage{bm}
\usepackage{enumitem}
\usepackage{color}
\definecolor{orange}{RGB}{255,127,0}

\newcommand{\bear}{\begin{eqnarray}}
\newcommand{\eear}{\end{eqnarray}}
\newcommand{\nn}{\nonumber}

\newcommand{\mF}{\mathcal{F}}

\newcommand{\mG}{\mathcal{G}}
\newcommand{\mH}{\mathcal{H}}

\newcommand{\mL}{\mathcal{L}}

\newcommand{\mO}{\mathcal{O}}

\newcommand{\cO}{{\cal O}}
\newcommand{\Frac}[2]{\frac{\displaystyle #1}{\displaystyle #2}}

\newcommand{\ket}{\,\rangle}
\newcommand{\bra}{\langle \,}
\newcommand\lsim{\mathrel{\rlap{\lower4pt\hbox{\hskip1pt$\sim$}}
    \raise1pt\hbox{$<$}}}
\newcommand\gsim{\mathrel{\rlap{\lower4pt\hbox{\hskip1pt$\sim$}}
    \raise1pt\hbox{$>$}}}
\newcommand{\ba}{\begin{array}}
\newcommand{\ea}{\end{array}}

\newcommand{\be}{\begin{equation}}
\newcommand{\ee}{\end{equation}}
\begin{document}
\title{Resonant production of $Wh$ and $Zh$ at the LHC}
%%%%%%%%%%%%%%%%%%%%%%%%%%%%%%%%%%%%%%%%%%%%%%%%%%%%%%%%%%%%%%%%%%%

\author{Antonio Dobado, Felipe J. Llanes-Estrada and Juan J. Sanz-Cillero     \\
{\it Departamento de F\'{\i}sica Te\'orica, Universidad Complutense de Madrid,}  \\
{\it Plaza de las Ciencias 1, 28040 Madrid, Spain}}

\maketitle

\begin{abstract}

We examine the production of $Wh$ and $Zh$ pairs
at the LHC in the context of a Strongly Interacting Symmetry Breaking Sector of the Standard Model.
Our description is based on a non-linear Higgs Effective Theory,
including only the Standard Model particles.
We focus on its scalar sector (Higgs boson $h$ and electroweak Goldstones associated to $W_L^\pm$ and $Z_L$), which is expected to give
the strongest beyond Standard Model  rescattering effects.
The range of the effective theory is extended with dispersion-relation based unitarization,
and compared to the alternative extension with explicit axial-vector resonances.
We estimate the $Wh$ and $Zh$ production cross-section,
where an intermediate axial-vector resonance is generated
for certain values of the chiral couplings.
We exemplify our analysis with a benchmark axial-vector with $M_A=3$~TeV.
Interestingly enough, these different approaches provide essentially
the same prediction.
Finally we discuss the sensitivity of ATLAS and CMS to such resonances.

\end{abstract}

\newpage

%%%%%%%%%%%%%%%%%%%%%%%%%%%%%%%%%%%%%%%%%%%%%%%%%%%%%%%%%%%%%%%%%%%%%
\section{Introduction}
%%%%%%%%%%%%%%%%%%%%%%%%%%%%%%%%%%%%%%%%%%%%%%%%%%%%%%%%%%%%%%%%%%%%%
If physics beyond the Standard Model exists, the absence of new physics states in the few-hundred GeV region
and the lightness of the new Higgs-like boson naturally suggest some new
strongly interacting sector with an additional (pseudo) Goldstone boson beyond
the three needed for the Electroweak Chiral Symmetry Breaking.
This would call for enlarging the Standard Model (SM) symmetry group, leading perhaps to Composite Higgs Models.

A complete exploration of the Goldstone boson scattering in those SM extensions
may provide important information on their underlying nature.
This requires not only an examination of $V_L V_L$ but also $V_L h$ amplitudes,
with $V=W^\pm,Z$.
For $M_{W,Z}\ll\sqrt{s}$, the scattering of longitudinal component $V_L$ of the massive electroweak (EW) gauge bosons is related to EW Goldstone ($\omega^a$) processes by the equivalence theorem (ET)~\cite{Cornwall:1974km,ETET}:
$T(V_L^a V_L^b\to V^c_L V^d_L) \simeq  T(\omega^a \omega^b\to \omega^c \omega^d)$,
$T(V_L^a h\to V^a_L h) \simeq  \, -\, T(\omega^a h\to \omega^a h)$.
We will extract the latter amplitude within the ET regime $M_{W,Z}\ll\sqrt{s}$
and neglect the $W^\pm$ and $Z$ masses.
Furthermore, since numerically $M_{q,\ell}\ll  M_{W,Z}\lsim M_h$ ($q\neq t$)~\footnote{
The top quark is not considered in this analysis and its impact in these scattering processes
via loops deserves a separate dedicated analysis. Nonetheless, some estimates
point out that these fermion corrections are subdominant~\cite{Castillo:2016erh,Delgado:2017cls}, as the scalar boson derivative interactions
win eventually over the non-derivative Yukawa contributions. },
all the masses of the remaining SM particles and their Yukawa couplings will also be neglected.

There are several  approaches to describing a modified SM including
a Strongly Interacting Symmetry Breaking Sector  (SISBS).
The existence of a large mass gap between the SM and possible new physics particles points out to effective field theories (EFT) as the most convenient and model independent
framework for the study of beyond SM (BSM) scenarios.
Likewise, we consider the non-linear representation of the EW Goldstones, as it provides
the most general SM extension allowed by symmetry~\cite{deFlorian:2016spz,Sanz-Cillero:2017jhb}.
Irrespective of the validity of those Composite Higgs Models (CHM),
the $\omega h$ scattering can be addressed in perturbation theory within
the Higgs Effective Field Theory
(HEFT)~\footnote{Not to be confused with a similarly named earlier theory that required the top to be much heavier than the Higgs and consisted of an expansion in inverse powers of $M_t$.
}~\cite{deFlorian:2016spz,EWChL}.
It describes the interactions of the would-be-Goldstone bosons $\omega^a$ from the spontaneous EW
symmetry breaking. Following the CCWZ formalism~\cite{CCWZ} the $\omega^a$
transform non-linearly under chiral transformations
and parametrize the coset $\mG/\mH$,
with the EW chiral group $\mG=SU(2)_L\times SU(2)_R$
and the global custodial vector subgroup $\mH=SU(2)_{R+L=C}$.
HEFT extends the Higgsless Electroweak Chiral Lagrangian~\cite{Higgsless-EWChL}
by the addition of one singlet scalar Higgs field $h$
with generic couplings~\cite{EWChL}; in this theory, one is agnostic about the nature of the Higgs,
which is coupled as the most general scalar boson that does not
disrupt the pattern of global electroweak symmetry breaking
$SU(2)_L\times SU(2)_R\to SU(2)_{L+R=C}$.
Chiral symmetry $\mG$   %%%=SU(2)_L\times SU(2)_R$
is spontaneously broken down to the global custodial vector subgroup $\mH$.

The subgroup $SU(2)_L \times U(1)_Y\subset \mG$ is then gauged, like in the SM, and this introduces the coupling to the transverse gauge bosons (and thus, to the quark constituents of the proton).
We will assume that the only source of custodial symmetry breaking is
the gauging of the $U(1)_Y$ group and that additional $SU(2)_{R+L}$ breaking operators
only appear at higher orders in the HEFT, being further suppressed.

In preparing this article we have kept in mind some recent hints in the $Vh$ searches
in ATLAS~\cite{ATLAS:2017ywd}, where a $3.3\sigma$ ($2.2\sigma$)
local (global) significance excess has been reported at $M_{Vh}\approx  3$~TeV.
However, this excess has not been confirmed by CMS~\cite{Sirunyan:2017wto},
so we do not commit to a fixed energy scale for any $Vh$ resonances.
Nonetheless, we consider it useful to show that such phenomenon can be
naturally described by the HEFT in the axial-vector $Vh$ channel with $IJ=11$ quantum numbers.
Thus, as a case of study, we will consider a benchmark scenario with
a resonance mass $M_A=3$~TeV and explore the feasibility of its search at the LHC.

The low-energy EFT is introduced in Sec.~\ref{sec:EFT}.
In Sec.~\ref{sec:pertamp}, we compute the most relevant LHC subprocess, $q\bar{q}' \to V_Lh$,
for the production of axial-vector resonances in the $V_Lh$ channel. Possible BSM effects are parametrized in an axial-vector form-factor $\mF_A(s)$ defined therein.
The strong $V_Lh$ rescattering in the HEFT, its unitarization and the generation of an axial-vector resonance are discussed in Sec.~\ref{sec:IAM}.
The low-energy form-factor $\mF_A(s)$ within the HEFT is computed in Sec.~\ref{sec:FFpert} and
its extension up to the resonance region is provided in Sec.~\ref{sec:FFres}.
For this, we consider models with alternative unitarization procedures or with explicit resonance Lagrangians,
all of them leading to identical conclusions.
In Sec.~\ref{sec:intermediateW} we perform a phenomenological analysis of the $W^\pm h$ production cross section at the LHC for the referred SISBS benchmark point, with an $M_A=3$~TeV resonance.
Finally, some concluding remarks are provided in Sec.~\ref{sec:Conclusions}.

%%%%%%%%%%%%%%%%%%%%%%%%%%%%%%%%%%%%%%%%%%%%%%%%%%%%%%%%%%%%%%%%%%%%%
\section{Low-energy EFT}\label{sec:EFT}
%%%%%%%%%%%%%%%%%%%%%%%%%%%%%%%%%%%%%%%%%%%%%%%%%%%%%%%%%%%%%%%%%%%%
\label{sec:EWChTFermions}

\subsection{Leading order Lagrangian}

At leading order, the Lagrangian of the scalar symmetry breaking sector (SBS), the modified SBS of the SM, is an $SU(2)_L \times U(1)_Y$ gauged $SU(2)_L\times SU(2)_R/SU(2)_C$
non-linear sigma model HEFT which includes the Higgs field $h$ as a singlet. Including chirally interacting fermions, the lowest order (LO) SISBS Lagrangian reads
\begin{eqnarray}
\label{LagrangianI}
{\cal  L}_{\rm LO}
= \frac{v^2}{4} F(h) Tr\left\{\left(D_\mu U\right)^\dagger D^\mu U\right\} %
      +\frac{1}{2}\partial_\mu h \partial^\mu h   \\      \nonumber
      -  V(h) + i\overline{Q}\gamma^\mu d_\mu Q
      - v G(h)\left[\bar{Q}_L^\prime UH_Q Q_R^\prime + \text{h.c.} \right],
\end{eqnarray}
where the matrix field  $U(\omega)\in SU(2)$
describes the EW would-be Goldstone fields (thus, by the ET, it gives us the $W_L$, $Z_L$ terms of the Lagrangian) and parametrizes
the coset $SU(2)_L\times SU(2)_R/SU(2)_C$.

The matrix $U$ is equivalently described by any unitary matrix that fulfills  $U=1 + i \sigma_a \omega^a/v + \cO(\omega^2)$.
In particular, in the spherical representation,
the would-be Goldstone fields $\omega^a$ are parametrized in the form
\begin{equation}
U \, = \, \sqrt{1-\frac{\omega^2}{v^2}}+i\frac{\bar{\omega}}{v},
\end{equation}
where  $\bar{\omega}=\sigma_i\omega^i$
and $v=0.246$~TeV the Higgs vacuum expectation value (vev).

The $SU(2)_L\times U(1)_Y$ covariant derivative is then given by
\begin{equation}
D_\mu U = \partial_\mu U-ig\frac{\sigma_i}{2} W_\mu^i U + ig^\prime U\frac{\sigma_3}{2} B_\mu .
\end{equation}
In turn, the Higgs potential and the functions $F(h)$ and $G(h)$
are taken to have an analytical expansion in powers of $h$ around $h=0$
\begin{eqnarray} \label{expansions}
F(h) \,&=&\,
1\,+\, \sum_{n=1}^\infty f_n \left(\frac{h}{v}\right)^n
\,=\, 1+ 2a \Frac{h}{v} + b \left(\Frac{h}{v}\right)^2 \, +\, \cO(h^3)\, ,
\nn\\
G(h) \, &=&\, 1\, +\, \sum_{n=1}^\infty g_n \left(\Frac{h}{v}\right)^n\, ,
\nn\\
  V(h)\, &=& \, v^4\;\sum_{n=2}^\infty V_n \left(\frac{h}{v}\right)^n\ .
\label{Higgspotential}
\end{eqnarray}
In the SM one has $a=1$, $b=a^2$, $g_1=1$, $V_2=V_3 = M_h^2/2v^2$, $V_4 = M_h^2/8v^2$
and $f_{n\geq 3}=g_{n\geq 2}=V_{n\geq 5} = 0$. Deviations from
these values imply new physics.

 Even though the low-energy parameters could be determined from the underlying theory if it was known, from a bottom-up approach the effective couplings are in principle independent from each other and must be extracted from experimental data.
Furthermore, one naturally expects that some parameters are larger than others as specific low-energy couplings are related to resonances with specific quantum numbers
in the underlying theory, in principle with different masses
and couplings~\cite{Pich:2015kwa,Pich:2016lew}.
But as a low energy theory of many models of interest (CHM, dilaton models, etc.),
we take the coefficients of the Higgs self-potential to scale
as powers of the Higgs mass so that they are negligible against the derivative couplings in the TeV region where resonances may be found ($s\gg M_h^2$), and thus we set $V(h)\simeq 0$ as in earlier work~\cite{Delgado:2013loa}. This approximation is consistent with our use of the Equivalence Theorem, $M_W^2\sim M_h^2 \ll s$.

In the TeV region, we see once more that all masses (especially the masses of the light quarks, most abundant in the proton) are negligible: $s\gg M_{\rm Fer}$.
Therefore, the Yukawa interactions in Eq.~(\ref{LagrangianI})
are in turn negligible, and thus we set $G(h)\simeq 0$ in this work. This means that the leading process producing a $V_Lh$ pair is the chain proceeding by an intermediate transverse gauge boson, and not the direct emission of a longitudinal one from the fermions.
As the scope of this work is to address $W_Lh$ couplings, the possible appearance of the resonances in fermionic channels is not treated and has been presented elsewhere~\cite{Castillo:2016erh}.

%%%%%%%%%%%%%%%%%%%%%%%%%%%%%%%%%%%%%%%%%%%%%%%%%%%%%%%%%%%%%%%%%%%%%%
\subsection{Next-to-leading order effective Lagrangian}
%%%%%%%%%%%%%%%%%%%%%%%%%%%%%%%%%%%%%%%%%%%%%%%%%%%%%%%%%%%%%%%%%%%%%%

At next-to-leading order (NLO), the relevant $\omega h$ interaction and production will be provided by the
Lagrangian~\cite{EWChL,Pich:2015kwa,Pich:2016lew,Delgado:2013hxa},
\begin{eqnarray}
\mL_{\rm NLO} & = & d \,
\frac{(\partial_\mu h\partial^\mu h)}{v^2} \,
{\rm Tr}\{ D_\nu U^\dagger D^\mu U\}
+e\, \frac{(\partial_\mu h\partial^\nu h)}{v^2} \,
{\rm Tr}\{ D^\mu U^\dagger D_\nu U\}
\nn\\
&&
- i f_9 \, \Frac{(\partial_\mu h)}{v} \, {\rm Tr}\{
\hat{W}^{\mu\nu} \, D_\nu U\, U^\dagger
- \hat{B}^{\mu\nu} \,U^\dagger \,D_\nu U \} \, ,
\label{eq:NLO-Lagr}
\end{eqnarray}
where we used for the field-strength tensors
the notation from~\cite{Pich:2016lew}:
\bear
\hat{W}_{\mu\nu} &=& \partial_\mu \hat{W}_\nu
- \partial_\nu \hat{W}_\mu - i [\hat{W}_\mu,\hat{W}_\nu]\, ,
\qquad
\hat{B}_{\mu\nu}= \partial_\mu \hat{B}_\nu - \partial_\nu\hat{B}_\mu
- i [\hat{B}_\mu , \hat{B}_\nu]\, ,
\nn\\
\hat{W}_\mu &=& -\Frac{g W^a_\mu \sigma_a}{2} \, ,
\qquad
\hat{B}_\mu = - \Frac{g' B_\mu \sigma_3}{2}\, .
\eear

As earlier in Eq.~(\ref{expansions}) all
the ``coefficients'' in front
of the Lagrangian operators are promoted to actual functions
of the Higgs field with an analytical expansion in powers of
$h$ around $h=0$. For example, $f_9$ is really the first term in an expansion $\mF_9(h)= f_9 +\cO(h)$~\cite{Pich:2015kwa,Pich:2016lew}, but the $\cO(h)$ terms will be unnecessary unless processes with several
Higgs bosons (or higher orders of perturbation theory) are addressed.

Just as we did for the Higgs potential $V(h)$,
in the analysis in this paper we will assume that
the counting of any NLO fermionic operators is suppressed by a power of the fermion mass.
That couplings of the new scalar (Higgs) boson to fermions are indeed proportional to their masses
is what phenomenological analysis seem to be suggesting, both directly from Higgs-related measurements~\cite{Khachatryan:2016vau} and from flavor-factory legacy.

%%%%%%%%%%%%%%%%%%%%%%%%%%%%%%%%%%%%%%%%%%%%%%%%%%%%%%%%%%%%%%%%%%%%%%%%%%%%%%%%%%%%%%%%%%%%%%%%%%%
\section{The elementary subprocesses \\ $q\overline{q'} \rightarrow W_L/Z_L+h$ at leading order}\label{sec:pertamp}
%%%%%%%%%%%%%%%%%%%%%%%%%%%%%%%%%%%%%%%%%%%%%%%%%%%%%%%%%%%%%%%%%%%%%%%%%%%%%%%%%%%%%%%%%%%%%%%%%%%

In this work we address  the resonant production of $W^\pm h$ or $Zh$ pairs at the LHC. At very high energies the corresponding cross sections are very small unless the spontaneous symmetry breaking of the SM is strongly interacting. In that case the longitudinal components of the electroweak gauge bosons dominate the production and may have large enough cross sections to be detectable at the LHC through the subprocesses appearing in the title of this section. For the angular momentum $J$ and custodial isospin $I$ both equal to one, the corresponding amplitudes will be estimated with the Feynman diagram in Figure 1.

\begin{figure}
\centerline{\includegraphics[width=3cm]{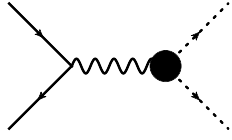}}
\caption{Tree-level Feynman diagram leading the production of $Wh$ and $Zh$ via the annihilation of  $q\bar{q'}$ quarks into a gauge $W^\pm (Z)$ boson. Strong rescattering in the final state appears through the form factor $\mF_A(s)$ represented by the thick blob.}
\label{fig:tree}
\end{figure}

In the limit when the light-quark Yukawas are negligible,
the amplitude in Fig.~\ref{fig:tree} factorizes into the tree-level
productions $q\overline{q'} \to W^* \rightarrow W_Lh$ and $q\overline{ q'}  \to Z^* \rightarrow Z_Lh $ and an axial form factor $\mF_A(s)$ encoding the strong rescattering $V_Lh$. For SISBS theories, this form factor is clearly
of a non-perturbative nature.
In this work, it will be computed by using the LO Electroweak Chiral Lagrangian of Eq.~(\ref{LagrangianI}) up to the one-loop level
and the NLO Lagrangian~(\ref{eq:NLO-Lagr}) at tree-level,
complemented with dispersion relations (unitarization of the amplitudes) and the Equivalence Theorem, as exposed next in sections~\ref{sec:FFpert} and~\ref{sec:FFres}.

At LO in the HEFT, the tree-level amplitudes of the quark-antiquark subprocesses  (thus, not including any form factor yet), are given, in their center of mass (CM), by

\begin{eqnarray}
T(u_-\bar{d}_+ \rightarrow W^+_Lh) & = &    \frac{g^2}{\sqrt{2}}  aV_{ud}\frac{\sqrt{s}E_W}{s-M_W^2} \sin\theta e^{-i\varphi}\\
T(d_-\bar{u}_+ \rightarrow W^-_Lh)  & = &    \frac{g^2}{\sqrt{2}}  aV_{ud}^*\frac{\sqrt{s}E_W}{s-M_W^2} \sin\theta e^{-i\varphi}\\
T(u_-\bar{u}_+ \rightarrow Z_Lh)    & = &    \frac{e^2}{2s^2_Wc^2_W}  a \alpha_u\frac{\sqrt{s}E_Z}{s-M_Z^2} \sin\theta e^{-i\varphi}\\
T(d_-\bar{d}_+ \rightarrow Z_Lh)    & = &    \frac{-e^2}{2s^2_Wc^2_W}  a \alpha_d\frac{\sqrt{s}E_Z}{s-M_Z^2} \sin\theta e^{-i\varphi}\\
T(u_+\bar{u}_- \rightarrow Z_Lh)    & = &    \frac{-e^2}{2s^2_Wc^2_W}  a \beta_u\frac{\sqrt{s}E_Z}{s-M_Z^2} \sin\theta e^{i\varphi}\\
T(d_+\bar{d}_- \rightarrow Z_Lh)    & = &    \frac{e^2}{2s^2_Wc^2_W}  a \beta_d\frac{\sqrt{s}E_Z}{s-M_Z^2} \sin\theta e^{i\varphi},
\end{eqnarray}
where the $+$ and $-$ $u$ and $d$ (anti) quark subindices denote their helicity state and $a$ is the first parameter of the $F(h)$ function appearing
in the SISBS Lagrangian of Eq.~(\ref{LagrangianI}):
$F(h)= 1 + 2 a h/v + b h^2/v^2 +\cO(h^3)$
(notice that in the SM  $a=1$, $b=a^2$;
a separation thereof signals strong interactions).
Furthermore,
$s_W$ and $c_W$ are respectively the sine and cosine of the Weinberg angle and
\begin{eqnarray}
\alpha_u & = & 1- \frac{4}{3} s_W^2  \\   \nonumber
\alpha_d & = & 1- \frac{2}{3} s_W^2  \\   \nonumber
\beta_u & = & - \frac{4}{3} s_W^2  \\   \nonumber
\beta_d & = & - \frac{2}{3} s_W^2  \ .  \\   \nonumber
\end{eqnarray}
Finally $E_W$ and $E_Z$ are the energies of the produced $W$ and $Z$  gauge bosons and
$\theta$ and $\varphi$ are the corresponding polar and azimuthal CM scattering angles, respectively.

At the high energies in which we are interested here $ \sqrt{s}\gg M_W, M_Z, M_h$, those amplitudes become:
\begin{eqnarray} \label{pertampsqq}
T(u_-\bar{d}_+ \rightarrow W^+_Lh) & = &    \frac{g^2}{2\sqrt{2}}  aV_{ud}\sin\theta e^{-i\varphi}\\
T(d_-\bar{u}_+ \rightarrow W^-_Lh)  & = &    \frac{g^2}{2\sqrt{2}}  aV_{ud}^* \sin\theta e^{-i\varphi}\\
T(u_-\bar{u}_+ \rightarrow Z_Lh)    & = &    \frac{g^2}{4c^2_W}  a \alpha_u \sin\theta e^{-i\varphi}\\
T(d_-\bar{d}_+ \rightarrow Z_Lh)    & = &    \frac{-g^2}{4c^2_W}  a \alpha_d \sin\theta e^{-i\varphi}\\
T(u_+\bar{u}_- \rightarrow Z_Lh)    & = &    \frac{-g^2}{4c^2_W}  a \beta_u\sin\theta e^{i\varphi}\\
T(d_+\bar{d}_- \rightarrow Z_Lh)    & = &    \frac{g^2}{4c^2_W}  a \beta_d \sin\theta e^{i\varphi} \ .
\end{eqnarray}

Guided by the precision LEP observables, we assume that custodial $SU(2)_{L+R}$ symmetry is a good approximation to the electroweak SISBS. This is obtained in the limit $g'=0$ (which implies $s_W=0$, $c_W=1$ so that $\alpha_u=\alpha_d=1$ and $\beta_u=\beta_d=0$).
As experimentally $|V_{ud}|\simeq 0.9758$, we will take $V_{ud}=1$.
In the following we will simplify the amplitudes of Eq.~(\ref{pertampsqq}) with that approximation. Then the non-vanishing ones are given by the simpler formulae
\begin{eqnarray} \label{pertamps}
T(u_-\bar{d}_+ \rightarrow W^+_Lh) & = &   T(d_-\bar{u}_+ \rightarrow W^-_Lh) = \frac{g^2}{2\sqrt{2}}  a \sin\theta e^{-i\varphi}\\
T(u_-\bar{u}_+ \rightarrow Z_Lh)    & = &  -T(d_-\bar{d}_+ \rightarrow Z_Lh)  = \frac{g^2}{4}  a  \sin\theta e^{-i\varphi}\,
\end{eqnarray}
whereas, because $\beta_{u/d}\to 0$, $T(u_+\bar{u}_- \rightarrow Z_Lh)=T(d_+\bar{d}_- \rightarrow Z_Lh)=0$.

In the presence of strong final state interactions, the amplitudes need to be modified by the introduction of an axial form factor $\mF_A(s)$. Thus the complete results will have the form
\begin{eqnarray}
\tilde  T(u_-\bar{d}_+ \rightarrow W^+_Lh)  = \tilde T(d_-\bar{u}_+ \rightarrow W^-_Lh) &=& \frac{g^2}{2\sqrt{2}}  a \sin\theta e^{-i\varphi}\mF_A(s)\ ,
\nn\\
\tilde  T(u_-\bar{u}_+ \rightarrow Z_Lh)  = \, - \, \tilde T(d_-\bar{d}_+ \rightarrow Z_Lh) &=& \frac{g^2}{4}  a \sin\theta e^{-i\varphi}\mF_A(s)\ .
\end{eqnarray}
The nonperturbative computation is thus isolated into computing the form factor $\mF_A(s)$.
Next, in section~\ref{sec:IAM} we study in detail the $W_Lh$ and $Z_Lh$ interactions and show how axial resonances arise out of these interactions.

%%%%%%%%%%%%%%%%%%%%%%%%%%%%%%%%%%%%%%%%%%%%%%%%%%%%%%%%%%%%%%%%%%%%%%%%%%%%%%%%%%%%%%%%%%%%%%%%%%%
\section{The strongly interacting $W_Lh$ and $Z_Lh$ amplitudes}\label{sec:IAM}
%%%%%%%%%%%%%%%%%%%%%%%%%%%%%%%%%%%%%%%%%%%%%%%%%%%%%%%%%%%%%%%%%%%%%%%%%%%%%%%%%%%%%%%%%%%%%%%%%%%
In order to obtain the axial form factor that dresses the amplitudes in Eq.~(\ref{pertamps}) one needs to have at hand an appropriate and as general as possible a description of elastic $W_Lh$ and $Z_Lh$ scattering. Our approach here will start from the effective  Electroweak Chiral Lagrangian in Eqs.~(\ref{LagrangianI}) and~(\ref{eq:NLO-Lagr}). Then we will use the Equivalence Theorem \cite{Cornwall:1974km} as applied to this kind of Lagrangian  \cite{ETET}. The theorem relates the electroweak amplitudes (in renormalizable gauges) involving longitudinal components of the $W$ and $Z$ gauge bosons with the  ones involving the corresponding would-be Goldstone bosons at high energies. In the case of interest here it reads (in the CM rest frame)
\begin{equation} \label{EqTh}
T\left(W_L ^\pm (Z_L) h \rightarrow W_L (Z_L)h\right) = \, -\, T\left(\omega^\pm (\omega^0)h \rightarrow \omega^\pm (\omega^0)h\right) + \mO\left(\frac{M_W}{\sqrt{s}}\right).
\end{equation}
Therefore at high energies we can have access to  the strongly interacting SBS of the SM by studying the elastic scattering of the longitudinal components of the $W$, $Z$ and the Higgs boson $h$.
While the gauge boson polarization is not yet systematically reconstructed at the LHC, it appears that it will soon become possible~\cite{Aguilar-Saavedra:2017zkn,Maina}.

The amplitude for the would-be Goldstone ($\omega$'s) bosons and $h$ can be computed at tree level from the Lagrangian in Eq.~(\ref{LagrangianI}) (order $s$). As this Lagrangian is not renormalizable, going to the one-loop level (oder $s^2$) requires the introduction of derivative counterterms depending on
new couplings, as is standard in chiral perturbation theory.
Upon renormalization, these couplings absorb the one-loop divergences of the elastic amplitudes and pararametrize,  in a systematic way, our ignorance about the underlying SISBS for these processes. Thus, up to NLO, the relevant scalar Lagrangian in spherical coordinates
is
\begin{eqnarray}
\label{bosonLagrangian}
{\cal L} & = & \frac{1}{2}\left(1 +2 a \frac{h}{v} +b\left(\frac{h}{v}\right)^2\right)
\partial_\mu \omega^a
\partial^\mu \omega^b\left(\delta_{ab}+\frac{\omega^a\omega^b}{v^2}\right)
\nonumber +\frac{1}{2}\partial_\mu h \partial^\mu h \nonumber  \\
 & + & \frac{4 a_4}{v^4}\partial_\mu \omega^a\partial_\nu \omega^a\partial^\mu \omega^b\partial^\nu \omega^b +
\frac{4 a_5}{v^4}\partial_\mu \omega^a\partial^\mu \omega^a\partial_\nu \omega^b\partial^\nu \omega^b  +\frac{g}{v^4} (\partial_\mu h \partial^\mu h )^2  \nonumber   \\
 & + & \frac{2 d}{v^4} \partial_\mu h\partial^\mu h\partial_\nu \omega^a  \partial^\nu\omega^a
+\frac{2 e}{v^4} \partial_\mu h\partial^\nu h\partial^\mu \omega^a \partial_\nu\omega^a
\end{eqnarray}
that we have described in detail in~\cite{Delgado:2015kxa}.  With this practical Lagrangian at hand we have computed the one-loop amplitudes for elastic processes involving Goldstone bosons and the Higgs.
In the present application we provide the amplitude $\omega h\to \omega h$ given by
\begin{equation}
T_{II_z}(\omega^{I_z} h \rightarrow \omega^{I'_z} h)=M(s,t,u)\delta_{I_zI'_z}
\end{equation}
where $I=1$ is the $SU(2)_{L+R}$ custodial isospin and $s$, $t$ and $u$ are the standard Mandelstam variables for massless particles since at high energies $(\sqrt{s}\gg M_h)$ we will be neglecting the Higgs (and vector boson) mass in agreement with Eq.~(\ref{EqTh}). Then we obtain
\begin{eqnarray} \label{invAmp}
M(s,t,u) & =  &\frac{a^2-b}{v^2}t+ \frac{2d^r(\mu)}{v^4}t^2+\frac{e^r(\mu)}{v^4}(s^2+u^2)\\   \nonumber
& + & \frac{a^2-b}{576\pi^2v^4}[(72-88a^2+16b+36(a^2-1)\log \frac{-t}{\mu^2}   \\  \nonumber
&+ & 3 (a^2-b       )    (\log \frac{-s}{\mu^2}+\log \frac{-u}{\mu^2}      ))t^2   \\   \nonumber
& + & (a^2-b       )    (26 - 9\log \frac{-s}{\mu^2}-3\log \frac{-u}{\mu^2}      ))s^2   \\   \nonumber
& +&  (a^2-b       )    (26 - 9\log \frac{-u}{\mu^2}-3\log \frac{-s}{\mu^2}      ))u^2 ]  \\   \nonumber
\end{eqnarray}
This can be obtained from our previously published~\cite{Delgado:2013hxa,Delgado:2014dxa,Delgado:2015kxa}  $\omega \omega \rightarrow hh$ amplitude by crossing. The renormalized  couplings $d^r(\mu)$ and $e^r(\mu)$ depend on the renormalization scale $\mu$ as
 \begin{eqnarray} \label{RGE}
d^r(\mu) & = & d^r(\mu_0) +\frac{1}{192 \pi^2}(a^2-b)\left[(a^2-b)-6(1-a^2)\right]  \log\frac{\mu^2}{\mu_0^2} \nonumber  \\
e^r(\mu)  & = & e^r(\mu_0)- \frac{1}{48 \pi^2}(a^2-b)^2 \log\frac{\mu^2}{\mu_0^2}\ .
\end{eqnarray}
so that the amplitude in Eq.~(\ref{invAmp}) is $\mu$ invariant.

If a resonance of definite spin $J$ appears dynamically or couples to $V_L h$ in any way, it should appear in the corresponding partial wave amplitudes. It is then convenient to compute the first few partial waves that dominate the amplitude of Eq.~(\ref{invAmp}) at low energy. The $I=J=1$ partial wave
needed for this work is given by
\begin{equation}
M_{11}(s)= \frac{1}{32\pi}\int_{-1}^1 x \, M(s,t,u) \, dx
\end{equation}
where $t=- s(1-x)/2$ and $u=-s(1+x)/2$. A direct computation of the integral shows that this partial wave adopts the generic form common to other scattering processes at NLO~\cite{Delgado:2013loa}:
\begin{equation} \label{PWA}
M_{11}(s)=M_{11}^{(0)}(s)+  M_{11}^{(1)}(s)       = K s + s^2
\left[ B(\mu)+ D \log \frac{s}{\mu^2}+ E \log \frac{-s}{\mu^2}   \right].
\end{equation}
where
\begin{eqnarray} \label{constsAmp}
K & = & \frac{a^2-b}{96 \pi v^2}    \\   \nonumber
B(\mu) & = & \frac{e^r(\mu)-2d(\mu)}{96 \pi v^4}  -\frac{a^2-b}{110592 \pi^3 v^4}\left(150(1-a^2)-83(a^2-b)\right)       \\   \nonumber
D & = & \frac{a^2-b}{4608 \pi^3 v^4}\left(3(1-a^2)-(a^2-b)\right)       \\   \nonumber
E & = & -\frac{(a^2-b)^2}{9216 \pi^3 v^4}      \    .
\end{eqnarray}
This axial-vector partial wave is defined in the whole complex $s$ plane and it has the expected left cut (LC) along the negative real $s$-axis and the unitarity or right cut (RC) along the positive real $s$-axis. The physical amplitude is obtained by taking $s=E_{\rm CM}^2+ i 0$, i.e. just over the RC, with $E_{\rm CM}$ being the total CM energy.

At low energies, phase space for channels with more particles suppresses inelastic amplitudes and
elastic unitarity on the physical region is rather well satisfied, so that
\begin{equation} \label{Unitarity1}
{\rm Im} M_{11}(s) = \mid  M_{11}(s) \mid^2 \ .
\end{equation}
However the NLO amplitude in Eq.~(\ref{PWA}) fulfills the unitarity condition at a perturbative level only,
\begin{equation}
{\rm Im} M_{11}^{(1)}(s) = \mid  M_{11}^{(0)}(s) \mid^2 \ .
\end{equation}
This is equivalent to the relation $E=-K^2/\pi$ among the constants of Eq.~(\ref{constsAmp}), which can be very easily checked.
The more demanding exact elastic unitarity condition of Eq.~(\ref{Unitarity1}) can be satisfied,
with only the NLO computation at hand, by using, among other  possibilities~\cite{Delgado:2015kxa}, the Inverse Amplitude Method (IAM~\cite{IAM}). According to it, the unitarized amplitude is given by
\begin{equation} \label{IAM}
\tilde M_{11}(s) = \frac{ M_{11}^{(0)}(s) ^2    }{ M_{11}^{(0)}(s)- M_{11}^{(1)}(s)}.
\end{equation}
This amplitude fulfills the exact elastic unitarity condition in Eq.~(\ref{Unitarity1}), it has the proper analytical structure (LC and RC), it is $\mu$-independent and its low-energy expansion coincides with the HEFT up to the NLO:
\begin{equation}
\tilde M_{11}(s)=M_{11}^{(0)}(s)+  M_{11}^{(1)}(s)+ O(s^3/v^6)\ .
\end{equation}
Moreover, for certain regions of the coupling space, this amplitude~(\ref{IAM})
can feature a pole at some point $s_0$ in the second Riemann sheet of the $s$ complex plane. Any such poles have a natural interpretation as dynamically generated resonances with mass $M$ and width $\Gamma$ given by the relation $s_0=M^2-i M \Gamma$. The IAM method has been extensively and successfully applied to ordinary Chiral Perturbation Theory to describe pion and kaon scattering and the associated resonances $f(500)$, $\rho$ and many others. Thus we may have some confidence that the method could work also in reproducing dynamical resonances in the context of the SISBS of the SM.

Since the IAM formula is compact and simply algebraic, as opposed to the difficult integral expressions of usual dispersion relations, we can study the position of its complex $s$-plane poles (resonances) directly.
In the case of axial resonances with mass $M_A$ and relative small width $\Gamma_A$ (so that $\gamma_A \equiv\Gamma_A/M_A \ll 1$ ) we find
\begin{eqnarray} \label{masaIAM}
M_A^2 & =  & \Frac{K}{B}\,=\,
v^2\frac{a^2-b}{e-2d+ \frac{a^2-b}{1152 \pi^2}[150(a^2-1)+83(a^2-b)]}    \\
\gamma_A & =  & \Frac{K^2}{B+D+E} \,=\, \frac{\gamma^0_A}{1-\frac{3}{\pi}\gamma_A^0(1+2\frac{a^2-1}{a^2-b})        }
\end{eqnarray}
where
\begin{equation} \label{anchuraIAM}
\Gamma_A^0 = M_A \gamma_A^0=\frac{a^2-b}{96 \pi v^2}M_A^3
\end{equation}
and the $d(\mu)$ and $e(\mu)$ couplings are evaluated at $\mu=M_A$ (it is necessary to state this since $d-2e$ is not, in general, renormalization-scale invariant). Obviously the region of parameters $a$, $b$, $d$ and $e$ yielding a dynamical axial resonance is defined as the region $M_A^2 >0$ (though below about 500 GeV our use of the Equivalence Theorem does not hold scrutiny anymore) and $\gamma_A > 0$.

Equation~(\ref{masaIAM}) shows that the LO parameters alone (that is, $d=e=0$ but $a^2-b \neq 0$) are sufficient to generate an axial resonance. This is generically broad, as the width is proportional to the same $(a^2-b)$ separation from the SM.
In the limit $b\to a^2$ fulfilled by dilatonic theories~\cite{dilaton} and the SM,
the axial-vector becomes narrow, $\gamma_A\to 0$, and gets a mass $M_A^2= 192\pi^2 v^2/[25(a^2-1)]$, implying $a>1$. Likewise, one gets the lower bound $M_A>3.6$~TeV for $a<1.16$. In the SM ($a\to 1$), this mass goes to infinity,
decoupling from the low-energy theory.
Such resonances are generically called ``dynamically generated'' and it is unclear whether they correspond to a new particle or field that should enter a fundamental Lagrangian, depending on how broad the width is. The textbook example of this behavior is the $f_0(500)$ or $\sigma$-meson in hadron physics.

On the other hand, the NLO coefficients $e$ or $d$ can yield a light resonance, as they suppress the numerator in Eq.(\ref{masaIAM}). The resonance is then narrower, as Eq.~(\ref{anchuraIAM}) shows a kind of KSFR relation: the width is proportional to the cube of the mass times a known combination of the coefficients that does not depend on $d$, $e$.  Very often one expects that a resonance dominated by the NLO Lagrangian terms is actually a physical particle, and there is ample work integrating out that high energy field from the underlying action to yield expressions for the EFT coefficients in terms of its properties.

\begin{figure}[h]
\centering
\includegraphics*[width=0.5\textwidth]{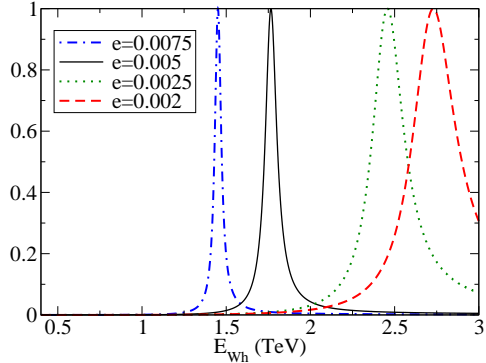}
\caption{The $I=J=1$ axial resonance generated in $V_L h$ scattering by the $e$ counterterm in the NLO HEFT Lagrangian, with values of the constant at $\mu=3$ TeV as indicated in the legend. Here, $d=0$, $a=0.95$ and $b=0.7 a^2$ are fixed. \label{fig:resonanceinA}}
\end{figure}

\begin{figure}[h]
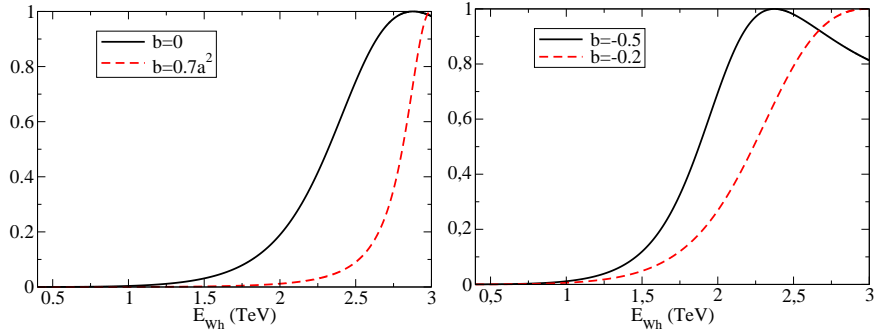

\centering
\includegraphics*[width=0.45\textwidth]{FIGS_DIR/WHres_deponb.eps}
\includegraphics*[width=0.45\textwidth]{FIGS_DIR/WHres_deponb2.eps}
\caption{The $I=J=1$ axial resonance generated in $V_L h$ scattering. Here we show the dependence on $b$, with $a$ fixed (on the left plot, to 0.95, on the right to -0.9) as well as fixing $e$ (to 1.64$\times 10^{-3}$ on the left plot, to 0 on the right). In both cases $d=0$.\label{fig:resonanceinA2}}
\end{figure}

\begin{figure}[h]
\centering
\includegraphics[width=0.5\textwidth]{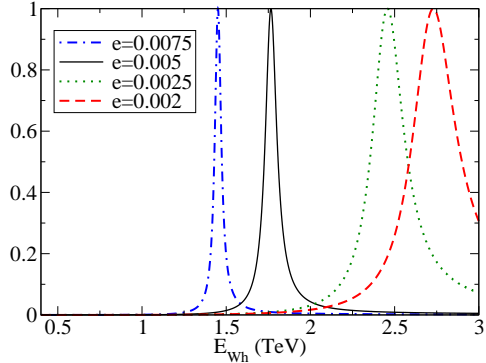}
\caption{The $I=J=1$ axial resonance generated in $V_L h$ scattering with $a=0.95$. Here we have fixed $2a^2-b=1$ which is a characteristic prediction of Minimally CHM. We set $d=0$. \label{fig:resonanceinA3}}
\end{figure}

Figures~\ref{fig:resonanceinA}, \ref{fig:resonanceinA2} and~\ref{fig:resonanceinA3} illustrate the resonances obtained in various cases of interest, depending on the Lagrangian parameters.
These amplitudes only depend on the combination $e(\mu)-2 d(\mu)$, so have fixed $d(\mu)=0$ and varied $e(\mu)$ in all the plots.

The method can accommodate a variety of resonances (or none).
Nevertheless, being based on an underlying Lagrangian, once its parameters are measured it does have predictive power yielding a specific spectrum and scattering amplitudes at higher energies.

%%%%%%%%%%%%%%%%%%%%%%%%%%%%%%%%%%%%%%%%%%%%%%%%%%%%%%%%%%%%%%%%%%%%%%%%%%%%%%%%%%%%%%%%%%%%%55
\section{The axial-vector form factor up to NLO in HEFT}\label{sec:FFpert}
%%%%%%%%%%%%%%%%%%%%%%%%%%%%%%%%%%%%%%%%%%%%%%%%%%%%%%%%%%%%%%%%%%%%%%%%%%%%%%%%%%%%%%%%%%%%%55
The piece connecting the Strongly Interacting Sector described in section~\ref{sec:IAM} with its perturbative coupling to the fermions of the Standard Model as in section~\ref{sec:pertamp} is the axial form factor $\mF_A(q^2)$ in the $\omega h$ sector that dresses the $V_L^*\to V_L h$ reaction. In this section we quickly compute it in perturbation theory, and defer the more sophisticated treatment necessary to address resonances for the next section.

In our treatment of the low-energy HEFT,  the necessary operators
at lowest order are provided by Eq.~(\ref{LagrangianI})
and those at next-to-leading order by Eq.~(\ref{eq:NLO-Lagr}). In particular,
in the ET limit the Axial Form Factor (AFF)  only depends on one NLO effective coupling, $f_9$.
At NLO in this limit this operator absorbs the ultraviolet divergences
cause by the one-lop AFF diagrams built out of the LO vertices from~(\ref{LagrangianI}).
In respecting custodial symmetry, the neutral-current form factor is provided by an isospin rotation and coincides with the charged one.

The computation of the $\omega h$ AFF $\mF_A(q^2)$ proceeds
by extracting the kinematic factors from the matrix element
\bear
\bra \omega^-(p_1) \, h(p_2)| J_{A}^\alpha |0\ket &=& (- i\sqrt{2}\, a)\, \mF_A(q^2) \,
\,P_T(q)^{\alpha\beta} (p_1-p_2)_\beta \, ,
\label{eq:AFF-def}
\eear
with $q=p_1+p_2$, $s=q^2$ in the timelike region for our application,
and the $L-R$ charged current being $J_{A}^\alpha= \frac{\delta S}{\delta a_\alpha}$
(with $a_\alpha= g W^+_\alpha/(2\sqrt{2})$ in the SM).
In practice this means that the vertex function for
$W^-_\alpha \to \omega^- h$ with external on-shell $\omega^-$ and $h$ (but $W^-$ off-shell)
is equal to
\bear
i \frac{a\, g}{2\sqrt{2} } \times (- i\sqrt{2})\,  \mF_A(s) \,
P_T(q)^{\alpha\beta} (p_1-p_2)_\beta
\, .
\label{eq:Wwh-vertex}
\eear

The normalization of the AFF defined in Eq.~(\ref{eq:AFF-def})
at zero momentum transfer is $\mF_A(0)=1$,
in consistency with the definition employed in previous sections.
To achieve this, a factor $a$ has been explicitly factorized out
(other works~\cite{Pich:2013fea,Pich:2012dv}
include this $a$ factor within $\mF_A(s)$ instead).

Within the ET and neglecting once more the Higgs and W and Z masses
at energies high enough over the $Wh$ threshold
one obtains the low-energy effective theory prediction
for the AFF up to NLO,
\begin{equation}
\mF_A(s)= \mF_A^{(0)}(s)+ \mF_A^{(1)}(s)+...
\end{equation}
where
\bear
\mF_A^{(0)}(s) &=& 1
\nn\\ \label{pertFF1}
 \mF_A^{(1)}(s)& = & s\, \left(G(\mu)+ H\ln \frac{-s}{\mu^2}  \right)    \, ,
\eear
in a notation analogous to Eq.~(\ref{PWA})
\begin{eqnarray}
G(\mu)&=&  -\, \frac{f_9(\mu)}{a \,v^2}
+\Frac{(a^2-b)}{36\pi^2 v^2} \, ,
\nn\\ \label{pertFF2}
H&=&  \,-\, \Frac{(a^2-b)}{96\pi^2 v^2}\, .
\end{eqnarray}

The NLO effective coupling $f_9$
renormalizes the one--loop divergence
(here in the $\overline{MS}$ scheme in dimensional regularization)
and runs with the scale in the form
 \begin{equation}
 f_9(\mu)= f_9(\mu_0)+\frac{a\,(a^2-b)}{96\pi^2}\log\frac{\mu^2}{\mu^2_0}
 \end{equation}
such that $\mF_A(s)$ is $\mu$ independent
and in agreement with the path integral
renormalization in~\cite{Guo:2015isa} (with notation $f_9=c_9(0)=\mF_9(0)$ therein).

The EFT result from Eq.~(\ref{pertFF1}) and~(\ref{pertFF2})
does not depend on whether the Goldstone field $U(\omega)$ is parametrized
in spherical, exponential, or any other coordinates.
This satisfying feature happens, in the ($s\gg M_h, M_W, M_Z\to 0$) approximation
because the four LO vertices active in the computation
($W\omega h$, $W\omega$, $h\omega\omega$ and $hh\omega\omega$) and the
one NLO vertex ($W\omega h$) all have at most two Goldstone fields each.

The AFF from Eq.~(\ref{pertFF1}) and~(\ref{pertFF2}) is an analytical function in the whole complex $s$-plane but for a RC, as expected.
On the other hand, in the elastic regime, unitarity relates
the imaginary part of the axial form factor with the partial-wave
scattering amplitude $M_{11}(s)$ in the form
\begin{equation} \label{Unitarity2}
{\rm Im} \mF_A(s) \,= \, \mF_A(s)\,  M_{11}(s)^*\, .
\end{equation}
However the one-loop result only fulfills this relation
at the perturbative level -i.e., up to NLO in the low-energy expansion--:
\begin{equation}
{\rm Im} \mF_A^{(1)}(s) \, =\,  \mF_A^{(0)}(s)\, M_{11}^{(0)}(s)^*
\,\,\, =\,\,  M_{11}^{(0)}(s),
\end{equation}
where on the last step we have used that $\mF_A^{(0)}(s)=1$ and
that the tree-level amplitude $M_{11}(s)$ is real.
This is easy to check comparing Eq.~(\ref{pertFF2}) and~(\ref{constsAmp}), which satisfy $H=-K/\pi$.
The reason of the violation of (exact) unitarity in the EFT calculation
is the absence of higher order corrections. As far as energies
remain small enough this deviations are negligible and our effective theory provides
an appropriate approximation of the physical amplitude.

%%%%%%%%%%%%%%%%%%%%%%%%%%%%%%%%%%%%%%%%%%%%%%%%%%%%%%%%%%%%%%%%%%%%%%%%%%%%%%%%%%%%%%%%%%%%%%%%%%%
\section{The Axial Form Factor in the resonance region}
\label{sec:FFres}
%%%%%%%%%%%%%%%%%%%%%%%%%%%%%%%%%%%%%%%%%%%%%%%%%%%%%%%%%%%%%%%%%%%%%%%%%%%%%%%%%%%%%%%%%%%%%%%%%%%

In this section we address the problem of how to obtain
an appropriate AFF $\mF_A(s)$ to describe
the $W_Lh$ and $Z_Lh$ resonant production at the LHC.
We deploy four different methods and show that, for relatively narrow resonances,
all give very similar results.

%%%%%%%%%%%%%%%%%%%%%%%%%%%%%%%%%%%%%%%%%%%%%%%%%%%%
\subsection{AFF with a resonance Lagrangian}
%%%%%%%%%%%%%%%%%%%%%%%%%%%%%%%%%%%%%%%%%%%%%%%%%%%%

The simplest approach, often employed by experimental collaborations in the search for new particles, is to include the resonance field explicitly as a degree of freedom in the Lagrangian~\cite{Pich:2016lew,Pich:2013fea}.
At tree-level, one finds a Breit-Wigner like formula
\bear
\mF_A(s) &=& 1 + \Frac{F_A\lambda_1^{hA}}{a\, v} \Frac{s}{M_A^2-s}
\nonumber \\  &=&  1 +   \Frac{s}{M_A^2-s}\, ,
\label{eq:axial-res-dom}
\eear
where the $F_A$ and $\lambda_1^{hA}$ constants
are respectively the $W\to A$ and $A\to \omega h$
vertex couplings; in the last identity, they are~\cite{Pich:2016lew,Pich:2013fea} fixed by
\bear
F_A\lambda_1^{hA}\,=\, a\, v
\eear
upon demanding that the AFF vanishes at asymptotically high energy (this depends on the underlying theory, and is typical, for example, of a non-Abelian gauge Lagrangian which yields asymptotic freedom).

Expanding  \ the AFF~(\ref{eq:axial-res-dom}) \ in powers \
of the squared four-momentum  $s$ one obtains the tree-level matching condition
$f_9 =  - F_A \lambda_1^{ha} v / M_A^2=  - a v^2/M_A^2$,
in agreement with previous works~\cite{Pich:2016lew}.

The intermediate resonance need not be infinitesimally narrow and its width $\Gamma_A$ can be
taken into account easily (which makes the form-factor regular on the real axis), yielding the relativistic Breit-Wigner line shape,
\begin{equation}
%\label{Formfactor1}
\mF_A(s) \, =\,  1\,  +\, \frac{s}{M_A^2-iM_A \Gamma_A-s}\, .  \hspace{3cm} ({\rm Model\ \ \  I})
\label{eq:res-lagr-model}
\end{equation}
In principle  $\Gamma_A$ is an independent parameter. Nevertheless, if the presumed $\omega h$ resonance is very elastic, suppressing other decay channels, the $\lambda_1^{ha}$ coupling of the resonance Lagrangian governs the width directly via
\bear
\Gamma_A &=& \Frac{\lambda_1^{hA} M_A^3}{48\pi v^2}\, \, =\,\,
\Frac{a^2 M_A^3}{48\pi F_A^2}\, ,
\eear
where the $W\to A$ coupling $F_A$ is expected to be of $\cO(v)$.

Adding the experimental constraints from the oblique $S$ and $T$ parameters,
further reduces the number of parameters.
For instance, under the assumption that the $W^3 B$ correlator obeys two Weinberg sum-rules
dominated by the lightest vector and axial-vector resonances~\cite{Pich:2013fea},
the axial-vector width becomes
\begin{equation} \label{Gammaconstrained}
\Gamma_A = a(1-a) \frac{M_A^3}{48\pi v^2}\ .
\end{equation}
Thus, for instance, a $M_A=3$~TeV resonance would have a width $\Gamma_A < 140$~GeV for
$0.95<a<1$.
A noticeable feature of Eq.~(\ref{Gammaconstrained}) is the absence of the ubiquitous factor $(a^2-b)$
--compare it, for example, with Eq.~(\ref{anchuraIAM})--. The reason is that the underlying effective Lagrangian including resonances explicitly correlates $a$ and $b$, so there is one less parameter.

Obviously, in less constrained scenarios where some of the previous theoretical assumptions are
relaxed,  one could obtain broader resonances.
But masses of a few TeV and widths of a few  hundred GeV naturally appear in HEFT frameworks if the  underlying theory is taken to be QCD-like.

 In the next subsection~\ref{subsec:unitFF}, we avoid introducing the resonance as an explicit degree of freedom affecting $\mF_A(s)$ and instead study it from
analyticity and unitarization of the low-energy HEFT amplitude.

%%%%%%%%%%%%%%%%%%%%%%%%%%%%%%%%%%%%%%%%%%%%%%%%%%%%%%%%%%%%%%%%%%%
\subsection{Unitarized HEFT parametrizations of the Axial Form Factor}
\label{subsec:unitFF}
%%%%%%%%%%%%%%%%%%%%%%%%%%%%%%%%%%%%%%%%%%%%%%%%%%%%%%%%%%%%%%%%%%%
Ideally, a fully realistic axial form factor $\mF_A(s)$ would have
the following properties:
\begin{enumerate}[label=\alph*]
\item Analiticity in the complex $s$ plane, featuring just a right cut for physical $s$. (We already know empirically that there are no bound state poles below threshold in the 100-Gev spectrum).   \label{itemAnal}\\
\item Coincidence of any resonance poles (in the second Riemann sheet) with those of the elastic amplitude $M_{11}(s)$. \label{itemPoles} \\
\item Elastic unitarity, i.e., $\mF_A$ should fulfill Eq.~(\ref{Unitarity2}),
while $M_{11}(s)$ satisfies  Eq.~(\ref{Unitarity1}).  \label{itemUnit} \\
\item Low-energy behavior that reproduces the chiral expansion $\mF_A(s)= \mF_A^{(0)}(s)+ \mF_A^{(1)}(s)+O(s^2/v^4)$.  \label{itemLow}\\
\end{enumerate}

Model I in Eq.~(\ref{eq:res-lagr-model}) features a resonance as in (\ref{itemPoles}) and can be matched to the low-energy expansion (\ref{itemLow}), but has no cut and bears no ressemblance to the elastic amplitude, so that (\ref{itemAnal}), (\ref{itemUnit}) and most of (\ref{itemPoles}) fail to be satisfied.

An alternative~\cite{IAM-FF} would be to build a form factor from a Lippman-Schwinger like resummation of the perturbative form factor expansion,
\begin{eqnarray}
\mF_A(s)\,&=&\, \frac{(\mF_A^{(0)}(s))^2}{\mF_A^{(0)}(s)-\mF_A^{(1)}(s)}
\,=\, 1 + \frac{\mF_A^{(1)}(s)}{1-\mF_A^{(1)}(s)} \nonumber
\\ \label{FormfactorF}
&=&\frac{1}{1-\mF_A^{(1)}(s)} \ ,  \hspace{3cm} ({\rm Model\ \ \  II})
\end{eqnarray}
which inherits from $\mF_A^{(1)}(s)$ the correct right cut, satisfying (\ref{itemAnal}) and, by construction, (\ref{itemLow}), but is again unconnected to the elastic amplitude, so it fails to fulfill (\ref{itemPoles}) and (\ref{itemUnit}).

From the elastic amplitude alone it is possible to build another form factor model~\cite{Dobado:1999xb,Espriu} that satisfies~(\ref{itemPoles}) and~(\ref{itemUnit})
\begin{eqnarray}
\mF_A(s) &=& 1+ \frac{ M_{11}^{(1)}(s)     }{ M_{11}^{(0)}(s)- M_{11}^{(1)}(s)} \nonumber \\
\label{Formfactor2}
&=& \frac{1}{1-\frac{M_{11}^{(1)}(s)}{M_{11}^{(0)}(s)}}
\ ,
\hspace{3cm} ({\rm Model\ \ \  III})
\end{eqnarray}
but, having no knowledge of $f_9$, which is in principle independent,  fails~(\ref{itemLow});
and since $M_{11}^{(1)}$ has a left cut, it fails also~(\ref{itemAnal}) (while this feature is probably not severe if employed in the resonance region only, from which that spurious left cut is very far in the complex $s$-plane).

One can improve Eq.~(\ref{Formfactor2}) by correcting for the low energy expansion, introducing $\mF_A^{(1)}$ as follows,
\begin{equation} \label{Formfactor4}
\mF_A(s) = 1+ \frac{\mF_A^{(1)}(s)   M_{11}^{(0)}(s)   }{ M_{11}^{(0)}(s)- M_{11}^{(1)}(s)}\ .
\hspace{3cm} ({\rm Model\ \ \  IV})
\end{equation}
This form factor satisfies all of~(\ref{itemPoles}), (\ref{itemUnit}) and~(\ref{itemLow}).
The only problem left is that, together with the RC, it has also a LC from $M_{11}^{(1)}(s)$.
But again, this LC is not expected to have a very strong influence in the physical timelike-$s$ region (the RC) particularly in the TeV, perhaps resonant, range of energies.

Interestingly, all three form factors in models II-IV, Eq.~(\ref{FormfactorF}), (\ref{Formfactor2}), (\ref{Formfactor4}) would coincide if
$ M_{11}^{(1)}(s) =\mF_A^{(1)}(s)   M_{11}^{(0)}(s) $.
This boils down to the following relations among the coefficients of $M_{11}^{(1)}(s)$ and those of $\mF_A^{(1)}(s)$,
\begin{eqnarray}
D & = & 0  \, ,
\label{eq:D=0}
\\
E & = &K H  \, ,
\label{eq:E=KH}
\\
B(\mu) & = & K G(\mu)  \, .
\label{eq:B=KG}
\end{eqnarray}
The first condition is equivalent to neglecting the LC contribution
and it is fulfilled for $b=a^2$ (as in the SM) or $b=4a^2-3$.
The second identity is always obeyed, since it is a consequence of perturbative unitarity.
The last one imposes a relation between  $f_9(\mu)$ and the rest of the couplings
so it can be fulfilled by a proper election of this parameter as a function of  those in the Lagrangian of Eq.~(\ref{eq:NLO-Lagr}),
namely $ a, \, b$ and the combination $e(\mu)-2d(\mu)$.

In general this choice of the parameter $f_9$ appears to be possible only at a given scale, so it would be $\mu$ dependent; however,
if Eqs.~(\ref{eq:D=0}) and~(\ref{eq:E=KH}) are obeyed then the right-hand and left-hand sides of
Eq.~(\ref{eq:B=KG}) have exactly the same running and all choices of $\mu$ are equivalent; the appropriate choice of $f_9$ is
\begin{equation} \label{eq:e-f9-relation}
\Frac{f_9}{a} \, =\, \frac{50(1-a^2)-17(a^2-b)}{384 \pi^2}+\frac{2d-e}{a^2-b}\ .
\end{equation}

\begin{figure}
\centering
\includegraphics[width=0.65\textwidth,clip]{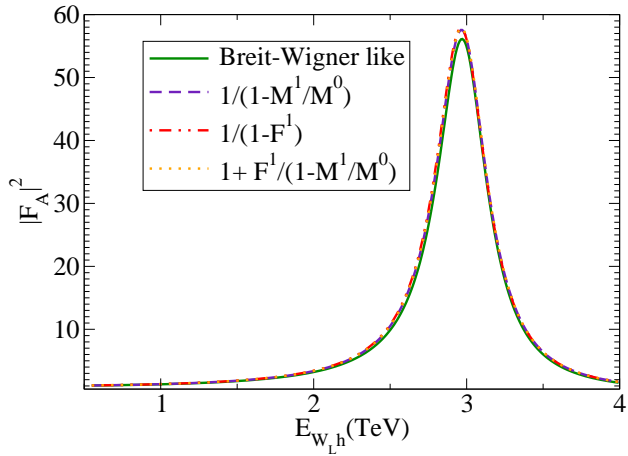}
\caption{\label{fig:FFs} The $I=J=1$ axial form factor in $V_L h$. Here we compare
various models of the form factor for fixed values of the chiral parameters. Because the resonance is relatively narrow, the form factor is controlled by its physical mass and width parameters, so the model differences are small (at the level of a percent). \label{fig:resonanceinFF}}
\end{figure}

\begin{table}
\caption{{\small Parameters employed to obtain the axial form factor of a relatively narrow $V_Lh$ resonance with mass around 3 TeV and width about 0.5 TeV, plotted in figure~\ref{fig:resonanceinFF}.\label{tab:params1}}}
\begin{tabular}{|ccc|c|}
\hline
& Model & Eq. in text & Parameters \\ \hline
I &  ($\mL_R$ Breit-Wigner like) %%%Resonance Lagrangian;
& (\ref{eq:res-lagr-model})&
$M_A,\, \Gamma_A$
%%%\\  & Breit-Wigner like) & &
\\ \hline
II & (Lipmann-Schwinger on pert. FF)& (\ref{FormfactorF})& $(a^2-b),\, f_9/a$
\\ \hline
III & (From elastic IAM only) & (\ref{Formfactor2})& $a,\, b,\, (e-2d)$ \\ \hline
IV & (Combined pert. FF + IAM) & (\ref{Formfactor4}) &  $a,\, b,\, f_9,\, (e-2d)$
\\ \hline
\end{tabular}
\end{table}

In figure~\ref{fig:resonanceinFF} we plot all four factor models $(I-IV)$ for a relatively narrow resonance in the neighbourhood of 3 TeV (this is achieved by appropriately setting either $f_9$ or $(e-2d)$, depending on the model, see table~\ref{tab:params1}). The agreement among them is spectacular (a consequence of the resonance being relatively narrow, so that the amplitude is pole-dominated) and therefore, it does not really matter what form factor model is used.

If one wants a quick cross-section estimate, the Breit-Wigner model (I) can as well be used; to use experimental data to constrain low-energy parameters of HEFT, the others should be implemented.

%%%%%%%%%%%%%%%%%%%%%%%%%%%%%%%%%%%%%%%%%%%%%%%%%%%%%%%%%%%%%%%%%%%%%a
\section{Cross section from intermediate gauge boson production}
\label{sec:intermediateW}
%%%%%%%%%%%%%%%%%%%%%%%%%%%%%%%%%%%%%%%%%%%%%%%%%%%%%%%%%%%%%%%%%%%%%
Now we are in  a condition to provide a quick estimate
for the resonant production of $W_h$ and $Z_h$ at the LHC. After this extensive discussion,
all pieces that enter the cross-section are at hand.
For example we have, for  the unpolarized CM cross-section,
\begin{eqnarray} \label{udbarWh-dif}
\Frac{d \hat{\sigma}(u \overline d \rightarrow W^+h)}{d \Omega_{CM}}&=& \frac{a^2}{64 \pi^2 s}
\left( \frac{1}{4} \right) \left( \frac{g^4}{8} \right) \mid \mF_A(s)\mid^  2 \sin ^2 \theta
\nonumber \\
&=&
a^2\, \Frac{1}{128\,s} \, \Frac{\alpha^2 }{s_W^4 }\, |\mF_A(s)|^2\, \sin^2\theta\, .
\end{eqnarray}
Integrating over the full solid angle, one obtains
\begin{eqnarray}
\label{udbarWh}
\hat{\sigma}(u \overline d \rightarrow W^+h) &=&
a^2\frac{\pi}{48s}\frac{\alpha^2}{s^4_W} \mid \mF_A(s)\mid^2\ .
\end{eqnarray}
The strongly interacting SBS dynamics is encoded in the form factor which can be resonant or not depending on the parameters of the effective Lagrangian.
If $CP$ is conserved by the SISBS (as in our HEFT calculation up to NLO),
the same formula provides $\sigma(d \overline u \rightarrow W^-_L h)$.
Likewise, the $\sigma(q\bar{q}\to Z_L h)$ production cross section
is given by Eqs.~(\ref{udbarWh-dif}) and (\ref{udbarWh})
times a multiplicative factor $(\alpha_q^2 +\beta_q^2)/2$,
and multiplied by the appropriate distribution functions and summed on $q=u,d$ for the production from $pp$ collisions.

\begin{figure}
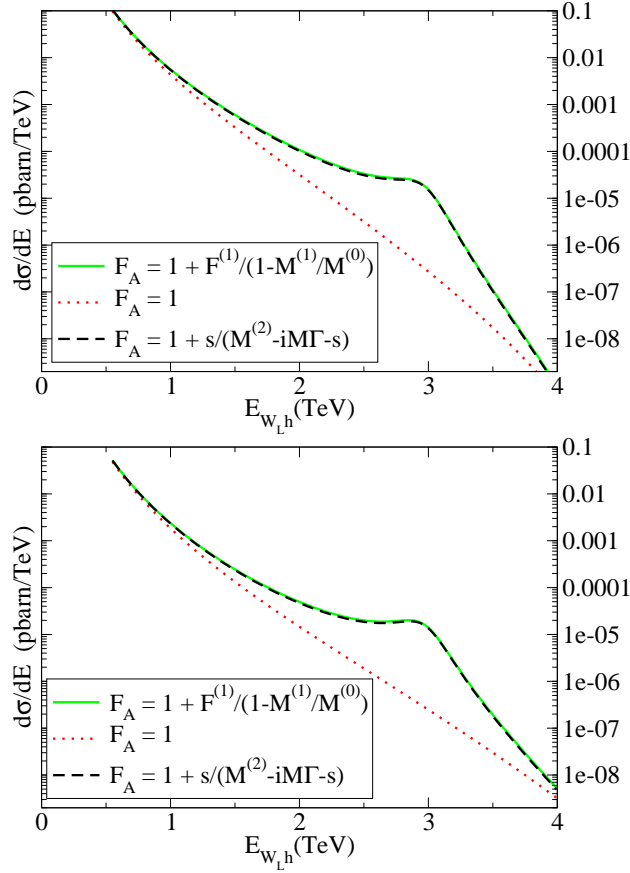

\centering
\includegraphics*[width=0.65\textwidth]{FIGS_DIR/CrossSecwh.eps}\\
\includegraphics*[width=0.65\textwidth]{FIGS_DIR/CrossSecwminush.eps}
\caption{\label{fig:Xsecwh} {\small Production cross section of $W_L^\pm  h$
%%%\simeq \omega h$
pairs with (top solid green and dashed black lines) and without (bottom dotted red line) a 3 TeV axial-vector resonance. The cross-section is enhanced by the latter, in this case by over an order of magnitude.
The top plot corresponds to $W^+h$ and the bottom one to $W^-h$; they are almost equal near the peak, with the positively charged one dominating for small $s_{Wh}$ and the negatively charged one at higher energy. Indeed, if we set the $\mF_A(s)$ form factor to 1
(dotted red lines),
the two cross-sections are very similar in the 3 TeV region.
}}
\end{figure}

Convoluting Eq.~(\ref{udbarWh}) with the parton distribution functions (which we take from the CJ (CTEQ-Jefferson Laboratory) set~\cite{Owens:2012bv}, that includes nuclear corrections, important at high $x$ and thus at the energy frontier of the LHC, as well as $Q^2$ corrections), we obtain the proton-proton level inclusive cross-section to produce a $W_Lh$ or $Z_Lh$  pair (by using the corresponding amplitudes given above) as:
\begin{eqnarray} \label{ppintermediateW}
\frac{d\sigma}{ds}(pp\to W_L^+ h+X) &=& \int^1_\frac{s}{E^2_{\rm tot}} \frac{dx_u}{x_u E^2_{\rm tot}} \hat{\sigma}_{u\bar{d}\to W_L^+ h} (s) F_{p/u}(x_u)
F_{p/\bar{d}}(x_{\bar{d}}) \nonumber
\\
\frac{d\sigma}{ds}(pp\to W_L^- h+X) &=& \int^1_\frac{s}{E^2_{\rm tot}} \frac{dx_d}{x_d E^2_{\rm tot}} \hat{\sigma}_{d\bar{u}\to W_L^- h} (s) F_{p/d}(x_d)
F_{p/\bar{u}}(x_{\bar{u}}) \ , \nonumber \\
\end{eqnarray}
with $x_{\bar{d}} = s/(x_uE^2_{\rm tot})$ and  $x_{\bar{u}} = s/(x_d E^2_{\rm tot})$.
Here, $s$ is the CM squared energy of the $Wh$ pair,
while $E_{\rm tot}$ is the CM energy of the $pp$ LHC accelerator.
A similar expression can be derived for the $Z_Lh$ production,
which provides a cross section of a similar order of magnitude
and will not be studied in this exploratory work.
For the example cross-section plotted in fig.~\ref{fig:Xsecwh}, we have set $E_{\rm tot}$ at 13 TeV.
There,  a resonance of mass 3 TeV and width 0.4 TeV has been injected with two of the form factors from fig.~\ref{fig:FFs}. The LO parameters are $a=0.95$, $b=0.7a^2$ (away from their SM values $a=b=1$), and the NLO ones $e(\mu)-2d(\mu)=1.64\times 10^{-3}$ and $f_9(\mu)=-0.6\times 10^{-2}$ for $\mu=3$~TeV.~\footnote{
This $f_9=-6\times 10^{-3}$, which leads to $M_A=3$~TeV and $\Gamma_A=0.4$~TeV, is very close to the value one would obtain from $e-2d$ through~(\ref{eq:e-f9-relation}), $f_9=-5.6\times 10^{-3}$.
The proximity of this two values relies on the fact that both expressions lead to the same resonance pole and
the conditions from~(\ref{eq:D=0})--(\ref{eq:B=KG}) for BSM theories, $b=4 a^2-3=0.61$, is approximately fulfilled
by our benchmark point $b=0.7 a^2=0.63$. }

Since the resonances here analyzed are native of the $W_Lh$ EW SBS, they are rather elastic and the branching fraction $R\to W_Lh$ is not too far below 1 and the difference therefrom can be ignored in a first experimental analysis (unlike other types of new physics that are weakly coupled to this channel).

We do find small cross-sections (fractions of a femtobarn) that are well below the current CMS and ATLAS cross-section upper bounds. The experimental collaborations are constraining $W'$ and $Z'$ models where the new resonance couples directly with charges $g_V=1$ and $g_V=3$, leading to femtobarn-size cross-sections.
On the other hand, our computations proceed by the diagram of fig.~(\ref{fig:tree}) with an intermediate $W_T$ gauge boson and are smaller  by a factor  $(g/g_{V})^4$.  This means that it will be arduous for the LHC to fully constrain the ``natural'' parameter space in the 3 TeV region. For this reason, we look forward to its high-luminosity upgrade.

%%%%%%%%%%%%%%%%%%%%%%%%%%%%%%%%%%%%%%%%%%%%%%%%%%%%%%%%%%%%%%%%%%%%%
\section{Conclusions}\label{sec:Conclusions}
%%%%%%%%%%%%%%%%%%%%%%%%%%%%%%%%%%%%%%%%%%%%%%%%%%%%%%%%%%%%%%%%%%%%%

The IAM %%%method
has been applied in this article to describe
the strong elastic $V_L h$ rescattering in the regime where the ET %%%Equivalence Theorem
applies ($s\gg M_W^2, M_Z^2, M_h^2$) that is, for energies above about 500 GeV,  and, at the same time, an EFT description in terms of the low energy degrees of freedom makes sense ($s\ll (4\pi f)^2$,
with $f$ the vev scale of spontaneous symmetry breaking in the strong BSM sector).

Notice again that we are neglecting masses and CKM mixing for this exploratory study.
Clearly, a more exhaustive analysis should take into
account the detector acceptance and appropriate kinematical cuts in the angular integration.
Likewise, the $W$ and $h$ are not directly detected, but rather their decay products. Nevertheless, this type of `realistic' analysis is out of the scope of this article
and is relegated to future studies.

%%%Generically, EFT-based methods cannot predict whether there is new physics within reach of the LHC.
%%%But
If the precision program of the LHC measures deviations from the SM in the low energy coefficients of the chiral Lagrangian (the relevant combinations of $a$, $b$, $d$ and $e$ for this work)
EFT-based approaches can predict whether there is new physics within reach of the LHC.
These methods are sufficiently robust to qualitatively predict whether there is a reasonable hope of detecting new physics resonances within the accelerator's energy reach, through equation~(\ref{masaIAM}). In our case, the axial-vector resonance mass and width and the low-energy parameters are constrained through the KSFR-like relation in Eq.~(\ref{anchuraIAM}).

If other unitarization methods such as the N/D or the (improved) K-matrix method are employed, the results are consistent, and the theoretical method-choice uncertainty is about 20\% in the determination of the resonance mass, as it was recently shown in~\cite{Delgado:2015kxa}. One generically expects this from any unitarization method that respects analyticity in the complex plane.
%
%%%It is a theorem of complex analysis that
To reproduce an analytic function in its entire domain of analyticity (for example, a scattering amplitude in the resonance region) it is enough to know it with enough precision in a finite segment
(for example, where the LHC can measure the low-energy coefficients)
and to provide an appropriate analytical extension.
Thus, unitary and analytic methods do have some predictive power. On the contrary, methods such as the old K-matrix  %%% from the 1950s,
are unitary but lack the right analytic structure, being less reliable.

Finally, it is worth remarking that our strongly interacting SBS analysis with chiral NLO couplings ($e-2d$ and $f_9$ here)
of the order of $10^{-3}$ leads to much smaller production cross sections than those currently tested by ATLAS and CMS~\cite{Aad:2015owa}. This naturally allows the presence of resonances with mass $M_A\lsim 3$~TeV,
while evading standing experimental bounds in $V h$ resonant production,  contrary to some of the theoretical models considered by the experimental collaborations.

%%%%%%%%%%%%%%%%%%%%%%%%%%%%%%%%%%%%%%%%%%%%%%%%%%%%%%%%%%%%%%%%%%%%%
\subsection*{Acknowledgments}
The authors thank Rafael L. Delgado for assistance and valuable comments at early stages of this investigation. We also want to thank M.J. Herrero and D. Espriu for useful comments and suggestions, and the members of UPARCOS for maintaining a stimulating intellectual atmosphere.
Work supported by Spanish grants MINECO:FPA2014-53375-C2-1-P, FPA2016-75654-C2-1-P  and the COST Action CA16108.

\medskip
%%%%%%%%%%%%%%%%%%%%%%%%%%%%%%%%%%%%%%%%%%%%%%%%%%%%%%%%%%%%%%%%%%%%%


\begin{thebibliography}{99}
%%%%%%%%%%%%%%%%%%%%%%%%%%%%%%%%%%%%%%%%%%%%%%%%%%%%%%%%%%%%%%%%%%%%%



%%%\bibitem{equivalence-theorem}
%\cite{Cornwall:1974km}
\bibitem{Cornwall:1974km}
  J.~M.~Cornwall, D.~N.~Levin and G.~Tiktopoulos,
  %``Derivation of Gauge Invariance from High-Energy Unitarity Bounds on the s Matrix,''
  Phys.\ Rev.\ D {\bf 10} (1974) 1145
   Erratum: [Phys.\ Rev.\ D {\bf 11} (1975) 972].
  doi:10.1103/PhysRevD.10.1145, 10.1103/PhysRevD.11.972;
  %%CITATION = doi:10.1103/PhysRevD.10.1145, 10.1103/PhysRevD.11.972;%%
  %1134 citations counted in INSPIRE as of 25 Oct 2017
%
%\cite{Vayonakis:1976vz}
%%%\bibitem{Vayonakis:1976vz}
  C.~E.~Vayonakis,
  %``Born Helicity Amplitudes and Cross-Sections in Nonabelian Gauge Theories,''
  Lett.\ Nuovo Cim.\  {\bf 17} (1976) 383.
  doi:10.1007/BF02746538
  %%CITATION = doi:10.1007/BF02746538;%%
  %334 citations counted in INSPIRE as of 25 Oct 2017
%
%\cite{Lee:1977eg}
%%%\bibitem{Lee:1977eg}
  B.~W.~Lee, C.~Quigg and H.~B.~Thacker,
  %``Weak Interactions at Very High-Energies: The Role of the Higgs Boson Mass,''
  Phys.\ Rev.\ D {\bf 16} (1977) 1519.
  doi:10.1103/PhysRevD.16.1519;
  %%CITATION = doi:10.1103/PhysRevD.16.1519;%%
  %1940 citations counted in INSPIRE as of 25 Oct 2017
%
%\cite{Chanowitz:1985hj}
%%%\bibitem{Chanowitz:1985hj}
  M.~S.~Chanowitz and M.~K.~Gaillard,
  %``The TeV Physics of Strongly Interacting W's and Z's,''
  Nucl.\ Phys.\ B {\bf 261} (1985) 379.
  doi:10.1016/0550-3213(85)90580-2;
  %%CITATION = doi:10.1016/0550-3213(85)90580-2;%%
  %1051 citations counted in INSPIRE as of 26 Oct 2017
%
%\cite{Gounaris:1986cr}
%%%\bibitem{Gounaris:1986cr}
  G.~J.~Gounaris, R.~Kogerler and H.~Neufeld,
  %``Relationship Between Longitudinally Polarized Vector Bosons and their Unphysical Scalar Partners,''
  Phys.\ Rev.\ D {\bf 34} (1986) 3257.
  doi:10.1103/PhysRevD.34.3257.
  %%CITATION = doi:10.1103/PhysRevD.34.3257;%%
  %270 citations counted in INSPIRE as of 25 Oct 2017
%%%%%%
%%%%%%
%%%%%%%\cite{Cornwall:1974km,Vayonakis:1976vz}
%%%%%%\bibitem{Cornwall:1974km}
%%%%%%  J.~M.~Cornwall, D.~N.~Levin and G.~Tiktopoulos,
%%%%%%  Phys.\ Rev.\ D {\bf 10}, 1145 (1974)
%%%%%%  [Phys.\ Rev.\ D {\bf 11}, 972 (1975)].
%%%%%%  C.~E.~Vayonakis,
%%%%%%  %``Born Helicity Amplitudes and Cross-Sections in Nonabelian Gauge Theories,''
%%%%%%  Lett.\ Nuovo Cim.\  {\bf 17}, 383 (1976). M.S. Chanowitz and M.K. Gaillard, Nucl. Phys. {\bf 261} (1985) 379.
%%%%%%  G.J. Gounaris, R. Kogerler and H. Neufeld, Phys.\ Rev.\ {\bf  D34} (1986) 3257.


\bibitem{ETET}
%%%\bibitem{SISBS-equivalence-theorem}
%\cite{Dobado:1993dg}
%%%\bibitem{Dobado:1993dg}
  A.~Dobado and J.~R.~Peláez,
  %``On The Equivalence theorem in the chiral perturbation theory description of the symmetry breaking sector of the standard model,''
  Nucl.\ Phys.\ B {\bf 425} (1994) 110
   Erratum: [Nucl.\ Phys.\ B {\bf 434} (1995) 475]
  doi:10.1016/0550-3213(94)90174-0, 10.1016/0550-3213(94)00533-K
  [hep-ph/9401202];
  %%CITATION = doi:10.1016/0550-3213(94)90174-0, 10.1016/0550-3213(94)00533-K;%%
  %68 citations counted in INSPIRE as of 25 Oct 2017
%
%\cite{Dobado:1994vr}
%%%\bibitem{Dobado:1994vr}
%%%  A.~Dobado and J.~R.~Pelaez,
  %``The Equivalence theorem for chiral lagrangians,''
  Phys.\ Lett.\ B {\bf 329} (1994) 469
   Addendum: [Phys.\ Lett.\ B {\bf 335} (1994) 554]
  doi:10.1016/0370-2693(94)90392-1, 10.1016/0370-2693(94)91092-8
  [hep-ph/9404239];
  %%CITATION = doi:10.1016/0370-2693(94)90392-1, 10.1016/0370-2693(94)91092-8;%%
  %60 citations counted in INSPIRE as of 25 Oct 2017
%
%\cite{GrosseKnetter:1994yp}
%%%\bibitem{GrosseKnetter:1994yp}
  C.~Grosse-Knetter and I.~Kuss,
  %``The Equivalence theorem and effective Lagrangians,''
  Z.\ Phys.\ C {\bf 66} (1995) 95
  doi:10.1007/BF01496584
  [hep-ph/9403291];
  %%CITATION = doi:10.1007/BF01496584;%%
  %24 citations counted in INSPIRE as of 25 Oct 2017
%
%\cite{He:1993qa}
%%%\bibitem{He:1993qa}
  H.~J.~He, Y.~P.~Kuang and X.~y.~Li,
  %``Proof of the equivalence theorem in the chiral Lagrangian formalism,''
  Phys.\ Lett.\ B {\bf 329} (1994) 278
  doi:10.1016/0370-2693(94)90772-2
  [hep-ph/9403283].
  %%CITATION = doi:10.1016/0370-2693(94)90772-2;%%
  %82 citations counted in INSPIRE as of 25 Oct 2017



%\cite{deFlorian:2016spz}
\bibitem{deFlorian:2016spz}
  D.~de Florian {\it et al.} [LHC Higgs Cross Section Working Group],
  %``Handbook of LHC Higgs Cross Sections: 4. Deciphering the Nature of the Higgs Sector,''
  doi:10.23731/CYRM-2017-002
  arXiv:1610.07922 [hep-ph].
  %%CITATION = doi:10.23731/CYRM-2017-002;%%

%\cite{Sanz-Cillero:2017jhb}
\bibitem{Sanz-Cillero:2017jhb}
  J.~J.~Sanz-Cillero,
  %``Resonances and loops: scale interplay in the Higgs effective field theory,''
  arXiv:1710.07611 [hep-ph].
  %%CITATION = ARXIV:1710.07611;%%


%\cite{Castillo:2016erh}
\bibitem{Castillo:2016erh}
  A.~Castillo, R.~L.~Delgado, A.~Dobado and F.~J.~Llanes-Estrada,
  %``Top–antitop production from $W^+_L W^-_L$ and $Z_L Z_L$ scattering under a strongly interacting symmetry-breaking sector,''
  Eur.\ Phys.\ J.\ C {\bf 77}, no. 7, 436 (2017)
  doi:10.1140/epjc/s10052-017-4991-6
  [arXiv:1607.01158 [hep-ph]].
  %%CITATION = doi:10.1140/epjc/s10052-017-4991-6;%%



%\cite{Delgado:2017cls}
\bibitem{Delgado:2017cls}
  R.~L.~Delgado, A.~Dobado, D.~Espriu, C.~Garcia-Garcia, M.~J.~Herrero, X.~Marcano and J.~J.~Sanz-Cillero,
  %``Production of vector resonances at the LHC via WZ-scattering: a unitarized EChL analysis,''
  arXiv:1707.04580 [hep-ph].
  %%CITATION = ARXIV:1707.04580;%%
  %4 citations counted in INSPIRE as of 25 Oct 2017

\bibitem{EWChL}
%\cite{Feruglio:1992wf}
%%%\bibitem{Feruglio:1992wf}
  F.~Feruglio,
  %``The Chiral approach to the electroweak interactions,''
  Int.\ J.\ Mod.\ Phys.\ A {\bf 8} (1993) 4937;
%%%  doi:10.1142/S0217751X93001946
%%%  [hep-ph/9301281].
  %%CITATION = doi:10.1142/S0217751X93001946;%%
  %125 citations counted in INSPIRE as of 20 Oct 2017
%
%\cite{Alonso:2012px}
%\bibitem{Alonso:2012px}
  R.~Alonso {\it et al.}, %%%, M.~B.~Gavela, L.~Merlo, S.~Rigolin and J.~Yepes,
  %``The Effective Chiral Lagrangian for a Light Dynamical "Higgs Particle",''
  Phys.\ Lett.\ B {\bf 722} (2013) 330
   Erratum: [Phys.\ Lett.\ B {\bf 726} (2013) 926];
%%%  doi:10.1016/j.physletb.2013.04.037, 10.1016/j.physletb.2013.09.028
%%%  [arXiv:1212.3305 [hep-ph]].
  %%CITATION = doi:10.1016/j.physletb.2013.04.037, 10.1016/j.physletb.2013.09.028;%%
  %120 citations counted in INSPIRE as of 19 Oct 2017
%
%\cite{Buchalla:2013rka}
%%%\bibitem{Buchalla:2013rka}
  G.~Buchalla, O.~Cat\`a and C.~Krause,
  %``Complete Electroweak Chiral Lagrangian with a Light Higgs at NLO,''
  Nucl.\ Phys.\ B {\bf 880} (2014) 552
   Erratum: [Nucl.\ Phys.\ B {\bf 913} (2016) 475].
%%%  doi:10.1016/j.nuclphysb.2016.09.010, 10.1016/j.nuclphysb.2014.01.018
%%%  [arXiv:1307.5017 [hep-ph]].
  %%CITATION = doi:10.1016/j.nuclphysb.2016.09.010, 10.1016/j.nuclphysb.2014.01.018;%%
  %106 citations counted in INSPIRE as of 19 Oct 2017





\bibitem{CCWZ}
%\cite{Coleman:1969sm}
%%%\bibitem{Coleman:1969sm}
  S.~R.~Coleman, J.~Wess and B.~Zumino,
  %``Structure of phenomenological Lagrangians. 1.,''
  Phys.\ Rev.\  {\bf 177} (1969) 2239.
  doi:10.1103/PhysRev.177.2239;
  %%CITATION = doi:10.1103/PhysRev.177.2239;%%
  %1773 citations counted in INSPIRE as of 25 Oct 2017
%
%\cite{Callan:1969sn}
%%%\bibitem{Callan:1969sn}
  C.~G.~Callan, Jr., S.~R.~Coleman, J.~Wess and B.~Zumino,
  %``Structure of phenomenological Lagrangians. 2.,''
  Phys.\ Rev.\  {\bf 177} (1969) 2247.
  doi:10.1103/PhysRev.177.2247.
  %%CITATION = doi:10.1103/PhysRev.177.2247;%%
  %1551 citations counted in INSPIRE as of 25 Oct 2017


\bibitem{Higgsless-EWChL}
    T. Appelquist and C. W. Bernard,
    Phys. Rev. D {\bf 22} (1980) 200;
%
    A. C. Longhitano,
    Phys. Rev. D {\bf 22} (1980) 1166;
%
    Nucl. Phys. B {\bf 188} (1981) 118.







%\cite{ATLAS:2017ywd}
\bibitem{ATLAS:2017ywd}
  The ATLAS collaboration [ATLAS Collaboration],
  %``Search for Heavy Resonances Decaying to a W or Z Boson and a Higgs Boson in the $q\bar{q}^{(\prime)}b\bar{b}$ Final State in $pp$ Collisions at $\sqrt{s}$ = 13 TeV with the ATLAS Detector,''
  ATLAS-CONF-2017-018.
  %%CITATION = ATLAS-CONF-2017-018;%%

%\cite{Sirunyan:2017wto}
\bibitem{Sirunyan:2017wto}
  A.~M.~Sirunyan {\it et al.} [CMS Collaboration],
  %``Search for heavy resonances that decay into a vector boson and a Higgs boson in hadronic final states at sqrt(s)=13 TeV,''
  arXiv:1707.01303 [hep-ex];
  %%CITATION = ARXIV:1707.01303;%%
  H.~Huang, on behalf of the CMS coll.,
  %``Searches for diboson resonances at CMS,''
  arXiv:1710.05230 [hep-ex];
  %%CITATION = ARXIV:1710.05230;%%
  A.~M.~Sirunyan {\it et al.} [CMS Collaboration],
  %``Combination of searches for heavy resonances decaying to WW, WZ, ZZ, WH, and ZH boson pairs in proton–proton collisions at $\sqrt{s}=8$ and 13 TeV,''
  Phys.\ Lett.\ B {\bf 774}, 533 (2017)
  doi:10.1016/j.physletb.2017.09.083
  [arXiv:1705.09171 [hep-ex]].
  %%CITATION = doi:10.1016/j.physletb.2017.09.083;%%

%\cite{Pich:2015kwa}
\bibitem{Pich:2015kwa}
  A.~Pich, I.~Rosell, J.~Santos and J.~J.~Sanz-Cillero,
  %``Low-energy signals of strongly-coupled electroweak symmetry-breaking scenarios,''
  Phys.\ Rev.\ D {\bf 93} (2016) no.5,  055041
  doi:10.1103/PhysRevD.93.055041
  [arXiv:1510.03114 [hep-ph]].
  %%CITATION = doi:10.1103/PhysRevD.93.055041;%%

%\cite{Pich:2016lew}
\bibitem{Pich:2016lew}
  A.~Pich, I.~Rosell, J.~Santos and J.~J.~Sanz-Cillero,
  %``Fingerprints of heavy scales in electroweak effective Lagrangians,''
  JHEP {\bf 1704} (2017) 012
  doi:10.1007/JHEP04(2017)012
  [arXiv:1609.06659 [hep-ph]].
  %%CITATION = doi:10.1007/JHEP04(2017)012;%%


\bibitem{Delgado:2013loa}
  R.~L.~Delgado, A.~Dobado and F.~J.~Llanes-Estrada,
  %``Light ‘Higgs’, yet strong interactions,''
  J.\ Phys.\ G {\bf 41} (2014) 025002
  [arXiv:1308.1629 [hep-ph]].



\bibitem{Delgado:2013hxa}
  R.~L.~Delgado, A.~Dobado and F.~J.~Llanes-Estrada,
  %``One-loop $W_LW_L$ and $Z_LZ_L$ scattering from the electroweak Chiral Lagrangian with a light Higgs-like scalar,''
  JHEP {\bf 1402} (2014) 121
  [arXiv:1311.5993 [hep-ph]].

%\cite{Delgado:2014dxa}
\bibitem{Delgado:2014dxa}
  R.~L.~Delgado, A.~Dobado and F.~J.~Llanes-Estrada,
  %``Possible new resonance from $W_L W_L$-$hh$ interchannel coupling,''
  Phys.\ Rev.\ Lett.\  {\bf 114}, no. 22, 221803 (2015)
  doi:10.1103/PhysRevLett.114.221803
  [arXiv:1408.1193 [hep-ph]].
  %%CITATION = doi:10.1103/PhysRevLett.114.221803;%%


%\cite{Khachatryan:2016vau}
\bibitem{Khachatryan:2016vau}
  G.~Aad {\it et al.} [ATLAS and CMS Collaborations],
  %``Measurements of the Higgs boson production and decay rates and constraints on its couplings from a combined ATLAS and CMS analysis of the LHC pp collision data at $ \sqrt{s}=7 $ and 8 TeV,''
  JHEP {\bf 1608}, 045 (2016)
  doi:10.1007/JHEP08(2016)045
  [arXiv:1606.02266 [hep-ex]].
  %%CITATION = doi:10.1007/JHEP08(2016)045;%%






%\cite{Aguilar-Saavedra:2017zkn}
\bibitem{Aguilar-Saavedra:2017zkn}
  J.~A.~Aguilar-Saavedra, J.~Bernabéu, V.~A.~Mitsou and A.~Segarra,
  %``The Z boson spin observables as messengers of new physics,''
  Eur.\ Phys.\ J.\ C {\bf 77}, no. 4, 234 (2017)
  doi:10.1140/epjc/s10052-017-4795-8

\bibitem{Maina}
E. Maina, contribution to the EPS-HEP international conference on high energy physics,
Venice, July 6th 2017, {\tt https://indico.cern.ch/event/466934/contributions/2575374/}



%\cite{Delgado:2015kxa}
\bibitem{Delgado:2015kxa}
  R.~L.~Delgado, A.~Dobado and F.~J.~Llanes-Estrada,
  %``Unitarity, analyticity, dispersion relations, and resonances in strongly interacting $W_LW_L$, $Z_LZ_L$, and hh scattering,''
  Phys.\ Rev.\ D {\bf 91}, no. 7, 075017 (2015)
  [arXiv:1502.04841 [hep-ph]].
  %%CITATION = ARXIV:1502.04841;%%



\bibitem{IAM}
%%%  T.N. Truong, Phys.\ Rev.\ Lett.\ {\bf 61} (1988) 2526;
  A. Dobado, M.J. Herrero and T.N. Truong,
  Phys.\ Lett.\ {\bf B235} (1990) 134;
  A. Dobado and J.R. Pelaez, Phys.\ Rev.\ {\bf D47} (1993) 4883;
  Phys.\ Rev.\ {\bf D} 56 (1997) 3057.




\bibitem{dilaton}
%\cite{Halyo:1991pc}
  E.~Halyo,
  %``Technidilaton or Higgs?,''
  Mod.\ Phys.\ Lett.\ A {\bf 8} (1993) 275.
  doi:10.1142/S0217732393000271;
  %%CITATION = doi:10.1142/S0217732393000271;%%
  %27 citations counted in INSPIRE as of 30 Oct 2017
%
%\cite{Goldberger:2008zz}
%\bibitem{Goldberger:2008zz}
  W.~D.~Goldberger, B.~Grinstein and W.~Skiba,
  %``Distinguishing the Higgs boson from the dilaton at the Large Hadron Collider,''
  Phys.\ Rev.\ Lett.\  {\bf 100} (2008) 111802
  doi:10.1103/PhysRevLett.100.111802
  [arXiv:0708.1463 [hep-ph]].
  %%CITATION = doi:10.1103/PhysRevLett.100.111802;%%
  %256 citations counted in INSPIRE as of 30 Oct 2017


%\cite{Pich:2013fea}
\bibitem{Pich:2013fea}
  A.~Pich, I.~Rosell and J.~J.~Sanz-Cillero,
  %``Oblique S and T Constraints on Electroweak Strongly-Coupled Models with a Light Higgs,''
  JHEP {\bf 1401} (2014) 157
  doi:10.1007/JHEP01(2014)157
  [arXiv:1310.3121 [hep-ph]].
  %%CITATION = doi:10.1007/JHEP01(2014)157;%%



%\cite{Pich:2012dv}
\bibitem{Pich:2012dv}
  A.~Pich, I.~Rosell and J.~J.~Sanz-Cillero,
  %``Viability of strongly-coupled scenarios with a light Higgs-like boson,''
  Phys.\ Rev.\ Lett.\  {\bf 110} (2013) 181801
  doi:10.1103/PhysRevLett.110.181801
  [arXiv:1212.6769 [hep-ph]].
  %%CITATION = doi:10.1103/PhysRevLett.110.181801;%%


%\cite{Guo:2015isa}
\bibitem{Guo:2015isa}
  F.~K.~Guo, P.~Ruiz-Femen\'\i a and J.~J.~Sanz-Cillero,
  %``One loop renormalization of the electroweak chiral Lagrangian with a light Higgs boson,''
  Phys.\ Rev.\ D {\bf 92} (2015) 074005
  doi:10.1103/PhysRevD.92.074005
  [arXiv:1506.04204 [hep-ph]].
  %%CITATION = doi:10.1103/PhysRevD.92.074005;%%


\bibitem{IAM-FF}
  T.N. Truong, Phys.\ Rev.\ Lett.\ {\bf 61} (1988) 2526.


\bibitem{Dobado:1999xb}
  A.~Dobado, M.~J.~Herrero, J.~R.~Pelaez and E.~Ruiz Morales,
  %``CERN LHC sensitivity to the resonance spectrum of a minimal strongly interacting electroweak symmetry breaking sector,''
  Phys.\ Rev.\ D {\bf 62}, 055011 (2000)
  [hep-ph/9912224].
  %%CITATION = HEP-PH/9912224;%%
  %61 citations counted in INSPIRE as of 06 Aug 2015









\bibitem{Espriu}
   D.~Espriu and F.~Mescia,
  %``Unitarity and causality constraints in composite Higgs models,''
  Phys.\ Rev.\ D {\bf 90}, no. 1, 015035 (2014)
  [arXiv:1403.7386 [hep-ph]].
  %%CITATION = ARXIV:1403.7386;%%
  %13 citations counted in INSPIRE as of 06 Aug 2015
D.~Espriu, F.~Mescia and B.~Yencho,
  %``Radiative corrections to WL WL scattering in composite Higgs models,''
  Phys.\ Rev.\ D {\bf 88}, 055002 (2013)
  [arXiv:1307.2400 [hep-ph]].
  %%CITATION = ARXIV:1307.2400;%%
  %19 citations counted in INSPIRE as of 06 Aug 2015
D.~Espriu and B.~Yencho,
  %``Longitudinal WW scattering in light of the ?Higgs boson? discovery,''
  Phys.\ Rev.\ D {\bf 87}, no. 5, 055017 (2013)
  [arXiv:1212.4158 [hep-ph]].
  %%CITATION = ARXIV:1212.4158;%%
  %27 citations counted in INSPIRE as of 06 Aug 2015





 %\cite{Owens:2012bv}
\bibitem{Owens:2012bv}
  J.~F.~Owens, A.~Accardi and W.~Melnitchouk,
  %``Global parton distributions with nuclear and finite-$Q^2$ corrections,''
  Phys.\ Rev.\ D {\bf 87}, no. 9, 094012 (2013)
  doi:10.1103/PhysRevD.87.094012
  [arXiv:1212.1702 [hep-ph]].
  %%CITATION = doi:10.1103/PhysRevD.87.094012;%%



%\cite{Aad:2015owa}
\bibitem{Aad:2015owa}
  G.~Aad {\it et al.} [ATLAS Collaboration],
  %``Search for high-mass diboson resonances with boson-tagged jets in proton-proton collisions at $\sqrt{s}$ = 8 TeV with the ATLAS detector,''
  arXiv:1506.00962 [hep-ex];
  %%CITATION = ARXIV:1506.00962;%%
  M.~Aaboud {\it et al.} [ATLAS Collaboration],
  %``Search for new resonances decaying to a $W$ or $Z$ boson and a Higgs boson in the $\ell^+ \ell^- b\bar b$, $\ell \nu b\bar b$, and $\nu\bar{\nu} b\bar b$ channels with $pp$ collisions at $\sqrt s = 13$ TeV with the ATLAS detector,''
  Phys.\ Lett.\ B {\bf 765}, 32 (2017)
  doi:10.1016/j.physletb.2016.11.045
  [arXiv:1607.05621 [hep-ex]].
  %%CITATION = doi:10.1016/j.physletb.2016.11.045;%%



\end{thebibliography}
\end{document}